\documentstyle[preprint,12pt,aps,tighten]{revtex}

\begin{document}


\title{CTMC calculations of electron capture and ionization in collisions of
  multiply charged ions with elliptical Rydberg atoms}
\author{J.~Lu$^{1}$, Z.~Roller-Lutz$^{2}$, H.O. ~Lutz$^{1}$}
\address{$^{1}$Fakult\"at f\"ur Physik, Universit\"at Bielefeld, 33501
Bielefeld, Germany}  
\address{$^{2}$Institute of Physics, Faculty of Medicine, Rijeka University,
    Rijeka, Croatia}   \maketitle

\begin{abstract}
We have performed classical trajectory Monte Carlo (CTMC) studies of electron
capture and ionization in multiply charged ($Q\leq 8$) ion-Rydberg atom
collisions at intermediate impact velocities. Impact parallel to the minor and
to the major axis, respectively, of the initial Kepler electron ellipse has
been investigated. The important role of the initial electron momentum
distribution found for singly charged ion impact is strongly disminished for
higher projectile charge, while the initial spatial distribution remains
important for all values of Q studied.\\

\end{abstract}










The use of coherent elliptical Rydberg states in ion-atom collision studies
(for recent papers on the subject cf. [1-3] and references therein) has not
only aided the intuitive understanding of the interaction dynamics , it also
illuminates the roles of the momentum and the spatial distributions of the
target electron states. In classical terms, the {\it momentum distribution}
can be widely varied simply by changing the eccentricity $\varepsilon$ of the
Rydberg ellipse without affecting the energy of the state. In particular, for
impact perpendicular to the major axis of the ellipse, the capture cross
section displays a maximum if ${\bf v_{p}}$ (the perihelion electron velocity) is
parallel and equal to the projectile velocity ${\bf v}$; this is believed to be due
to the matching electron momenta in the initial target and the final
projectile state. In contrast, the role of the {\it spatial distribution}
becomes most clearly visible if the impact velocity vector in adjusted
perpendicular to the minor axis of the Rydberg ellipse; in this case, the
electrons can be located either between the approaching ion and the target
nucleus ("upstream geometry") or behind the target nucleus, as seen from the
projectile ("downstream geometry") without otherwise changing the 
momentum distribution of the Rydberg state (i.e., its angular momentum $l$
and the principal quantum number $n$). The capture cross section in both cases
turns out to be quite different: it is much larger in the upstream case as
compared to the downstream case; apparently, in the corresponding region of
parameter space the spatial characteristics of the initial state determine the
outcome of the collision. These investigations have so far been restricted to
collisions with singly charged ions. Recently, however, it has become possible
to employ such targets in studies involving multiply charged ions [1,4]. In
another context (electron capture by multiply charged ion in the presence of
an external magnetic field) we have found indications [5] that for increasing
projectile charge Q the distortion of the initial state increasingly dominates
over the influence of different target electron distributions; we have
therefore performed an exploratory study of such systems which is the topic of
this Letter. 

We employ the classical trajectory Monte Carlo (CTMC) method which is quite
useful in particular for the description of quantum mechanically complex
systems, giving a good qualitative and often fairly quantitative agreement
with experimental data ( for more recent applications to the study of Rydberg
atom collisions cf., e.g., [1-3]). Structureless ions of charge Q between 1 and
8 collide with Rydberg target atoms with nuclear charge q=1 and principal
quantum number $n=25$. The geometry is chosen such that the direction of impact
is perpendicular to the angular momentum direction of the Kepler
ellipse. Specifically, two cases are studied: (i) the impact is parallel to
the minor axis, thus allowing to study the velocity matching phenomenon, and
(ii) impact parallel to the major axis, showing the effect of the spatial
orientation of the target electron ("upstream-downstream asymmetry"). A useful
quantity characterizing the electron orbit is the (generalized) eccentricity
$\varepsilon=\pm \sqrt{1-(l/n)^{2}}$. In case (i), the + sign identifies
orbits with the perihelion velocity $\bf{v_{p}}$ parallel (-, antiparallel) to
$\bf{v}$; in case (ii) it characterizes the upstream (-, downstream) geometry. The
impact velocity $v$ is scaled by $1/n$, the velocity of a circular Rydberg
state, i.e., $V=vn=1$ in this case; the number of MC cycles was adjusted
to obtain statistical uncertainties of less than 5\%. Care has been taken to
assure that the projectile starts sufficiently far from the target
(approximately $3.5 \times10^5$ atomic units a.u.) to correctly describe the
initial part of the trajectory; in view of the long range Coulomb potential and
the known sensitivity of Rydberg states to $l, m$ changing processes, this is
critical particularly for higher projectile charges Q (see also below).\\

\noindent {\it (i) Impact parallel to the minor axis}\\
Figure 1 shows the eccentricity-dependent charge capture cross section $\sigma
/Q$ for $V=1.66$ and Q ranging from 1 to 8. Velocity matching is obtained at
$\varepsilon = +0.47$; at low Q, the cross section displays the well-known
behavior with a pronounced maximum at this eccentricity and a deep minimum for
negative eccentricities which characterize a strongly elongated Kepler ellipse
with $\bf{v}$ antiparallel to $\bf{v_{p}}$. For increasing Q, this structure
is soon washed out; while it is still noticeable for Q=2 and 4, almost any
trace of the minimum has disappeared for Q=8. Inspection of electron
trajectories during the approach of the projectile ion reveals the reason: as
expected, the long range Coulomb force distorts the original Kepler ellipse
already at quite long distances. This distortion is quite regular, and reminds
of a Stark effect. Indeed, a simple estimate confirms this: for Q=8, an
electric field strength of 5V/cm (i.e., of the order of the fields applied to
the collision region in the experiments [1]) is attained at approximately
$10^{5}$ a.u. This initial state effect might be reduced in the experiment by
applying a strong field in the target region; however, this results also
illuminates an inherent weakness of the CTMC-approach: the slow rise of the
electric field may "in reality" induce adiabatic transitions between the many
Rydberg states which would be populated differently in the classical
calculation. This distortion of the initial state becomes quite severe at
distances below $10^{4}$ a.u., i.e., corresponding to several revolution of
the Rydberg electron about its nucleus; therefore, it is to be believed that
the washing out of the cross section structure is indeed a real
effect. Finally, we may add that also the impact parameter dependence of the
capture probability reflects the signature of this effect. While for Q=1 and
$\varepsilon = +0.47$ (the velocity matching situation) the capture
probability is rather concentrated about the perihelion position [6], it is
nearly symmetric about b=0 (the position of the target nucleus) for Q=8.\\

\noindent {\it (ii) Impact parallel to the major axis}\\
For this study, we choose $\varepsilon = 0.96$, corresponding to $l=7$. This
creates a quite elongated state which is not too non-classical (low
$l$). Figure 2 shows for Q=8 the cross sections for upstream ($\sigma _{u}$)
and downstream ($\sigma _{d}$) geometry, respectively, including the
contributions of the various "swaps" to the capture cross sections. A swap has
been defined as a passage of the electron through the midplane between
projectile and the target nucleus; note, however, that for asymmetric
collisions ($q \ne Q$ as studied here), this plane has
to cut the connection line between target and projectile nucleus at the saddle
point of the two respective Coulomb potentials (i.e, at a distance
$R/(1+\sqrt{q/Q})$ from the projectile, with R the distance of both nuclei). In
view of the discussion in section (i) above, the strong upstream-downstream
asymmetry seen in these data is at first glance somewhat surprising. Again,
inspection of the electron trajectories sheds light on this point: the slowly
increasing electric field of the approaching projectile causes a distortion
and precession of the Kepler ellipse, in general not strong enough, however, to revert
the upstream into a downstream geometry and {\it vice versa}. Therefore, upon
approach of the projectile into the actual close interaction the electron is
still mainly fore or aft, respectively, of the target nucleus, thus
qualitatively preserving the role of the initial spatial distribution of the
electron; interestingly, even 3-swap and higher swap processes can still be
discerned. The qualitatively different behavior of $\sigma _{u}$ and $\sigma _{d}$ beyond
$V=1.5$ is associated with differently rising cross sections for ionization;
in case of the downstream geometry it sets in at considerably smaller V-values as
compared to the upstream geometry. This is further clarified by the respective
impact parameter dependencies (Fig.3a,b): In the upstream situation, charge
exchange extends out to fairly large distances and ionization is still weak;
in the downstream situation, the maximum charge exchange probability is of the
same order as in the upstream case, however, it is limited to much smaller
impact parameters, and ionization is already quite strong.

To conclude, our analysis shows that for impact of multiply charged ions of
intermediate velocities the role of the initial electron momentum distribution
becomes weaker for increasing projectile charge. This is due to the strong
perturbation of the initial state by the approaching ion which induces
pronounced changes in the momentum distribution long before the actual close
interaction occurs. In contrast, the initial spatial orientation of the
electron continues to be important for all Q-values studied here. \\

\noindent {\bf Acknowledgment}\\ \noindent 
 This  work has  been
supported by the Deutsche Forschungsgemeinschaft (DFG).

\newpage

{\bf{References:}}\\

\begin{itemize}

\item [[1]]~\parbox[t]{13cm} {J. C.~Day, B. D.~DePaola, T.~Ehrenreich,
S. B.~Hansen, E.~Horsdal-Pedersen, Y.~Leontiev and K. S.~Mogensen 1997 {\it
Phys. Rev} A{\bf 56}, 4700;}

\item [[2]]~\parbox[t]{13cm} {D. M.~Homan, O. P.~Makarov, O. P.~Sorokina,
K. B.~MacAdam, M. F. V.~Lundsgaard, C. D.~Lin, N.~Toshima 1998, {\it Phys.Rev}
A{\bf 58}, 4565 ;}

\item [[3]]~\parbox[t]{13cm} {L.~Kristensen, T.~Bov$\acute{e}$,
B. D.~DePaola, T.~Ehrenreich, E.~Horsdal-Perdersen and O. E.~Povlsen
 {\it submitted};}

\item [[4]]~\parbox[t]{13cm} {B. D.~DePaola, {\it private communiation};}

\item [[5]]~\parbox[t]{13cm} {J.~Lu, S.~Bradenbrink, Z.~Roller-Lutz,
and H.O.~Lutz, 1999 {\it J.Phys. B: At.Mol.Opt.Phys}.{\bf 32}, L681;}

\item [[6]]~\parbox[t]{13cm} {S.~Bradenbrink, H.~Reihl, Z.~Roller-Lutz,
and H.O.~Lutz, 1995 {\it J.Phys. B: At.Mol.Opt.Phys}.{\bf 28}, L133;}

\end{itemize}













\newpage

\noindent {\bf{Figure caption}}

\begin{itemize}

\item[Fig. 1] Eccentricity-dependent capture cross section $\sigma/Q$ for
impact velocity $V=1.66$ (in units of the circular $n=25$ Rydberg electron
velocity) and different projectile charges Q. For the initial state, velocity
matching is obtained at $\varepsilon=+0.47$.

\item[Fig. 2] Charge capture cross sections for (a)upstream $\sigma_{u}$ and
(b)downstream $\sigma_{d}$ geometry; the eccentricity $\varepsilon = \pm
0.96$. The respective ionization cross sections are also given.

\item[Fig. 3] Impact parameter dependent probabilities of capture and
  ionization: (a) upstream geometry,  (b) downstream geometry. Impact
  parameter $b$ in atomic units; impact velocity $V=1.5$; projectile charge
  $Q=8$. 

\end{itemize}

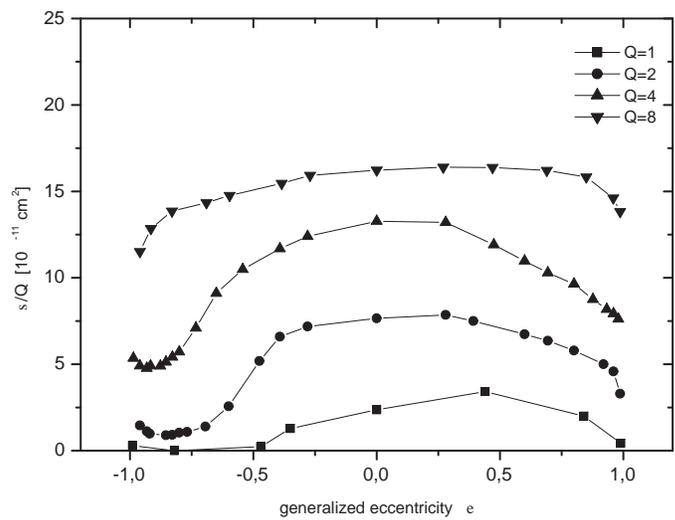
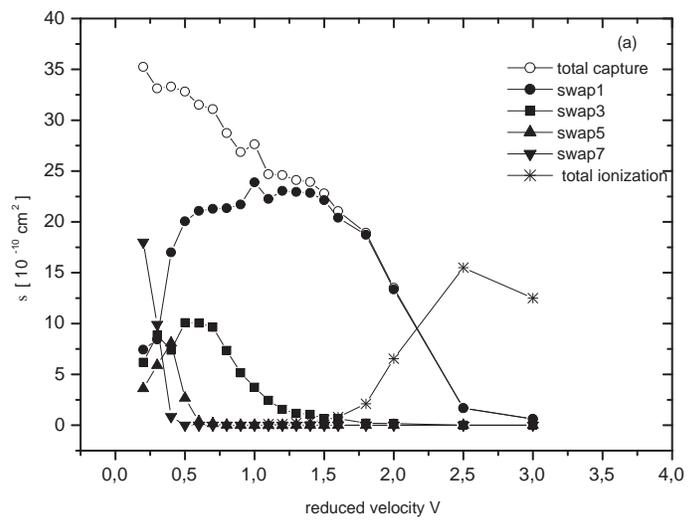
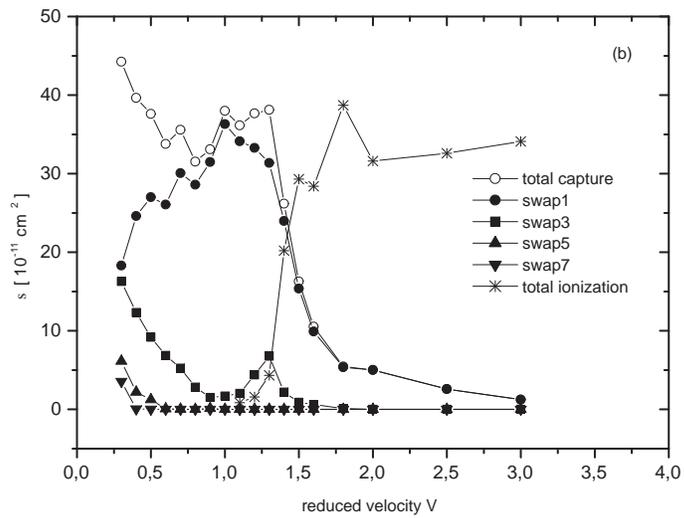
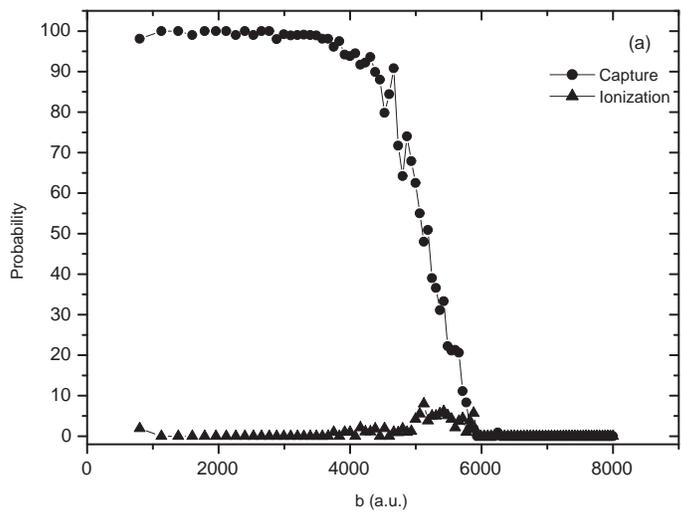
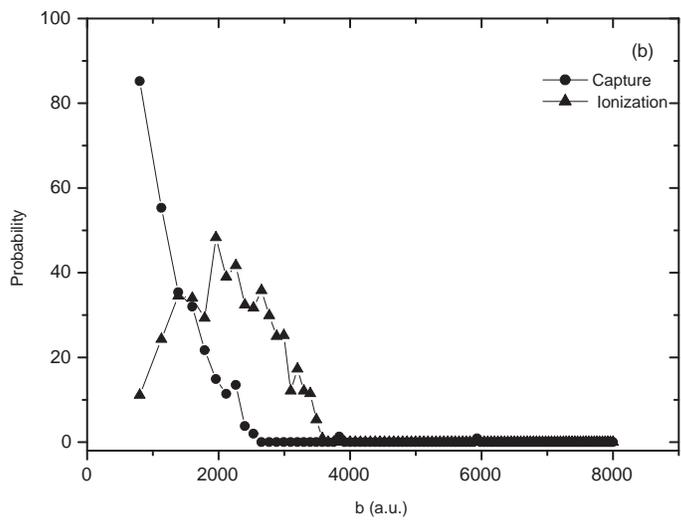
\end{document}